\documentclass[final,5p,times,twocolumn]{elsarticle}

\usepackage{slashed}
\usepackage{mathrsfs}
\usepackage{amsfonts}
\usepackage{amsmath}
\usepackage{amssymb}
\usepackage{revsymb}
\usepackage{graphicx,accents}
\usepackage{mathrsfs}
\usepackage{bm}
\usepackage{psfrag}
\usepackage{color}
\usepackage{hyperref}
\usepackage{natbib}
\usepackage{bbm}
\usepackage{bbold}
\usepackage{subcaption}
\usepackage{tabularx}

\usepackage[usenames,dvipsnames]{xcolor}

\newcommand{\be}{\begin{equation}}
\newcommand{\ee}{\end{equation}}
\newcommand{\ba}{\begin{array}}
\newcommand{\ea}{\end{array}}
\newcommand{\bea}{\begin{eqnarray}} 
\newcommand{\eea}{\end{eqnarray}} 
\newcommand{\bd}{\begin{displaymath}}
\newcommand{\ed}{\end{displaymath}}
\newcommand{\eps}{\varepsilon}

\newcommand{\epst}{\tilde{\varepsilon}}

\newcommand{\trm}[1]{\textrm{#1}}

\newcommand{\figref}[1]{Fig. \ref{#1}}
\newcommand{\figrefa}[1]{Fig. \ref{#1}(a)}
\newcommand{\figrefb}[1]{Fig. \ref{#1}(b)}

\newcommand{\eqnref}[1]{Eq. (\ref{#1})}
\newcommand{\eqnrefs}[2]{Eqs. (\ref{#1}) and (\ref{#2})}

\newcommand{\tr}{\trm{tr}\,}

\newcommand{\tsf}[1]{\textsf{#1}}

\newcommand{\vkap}{\varkappa}
\newcommand{\vphi}{\varphi}

\newcommand{\telta}{\tilde{\delta}}

\newcommand{\Sfi}{\tsf{S}_{f\!i}}

\newcommand{\J}{\trm{J}}

\newcommand{\nn}{\nonumber}

\newcommand{\e}{\mathbb{e}}
\newcommand{\gae}{g_{\phi e}}
\newcommand{\gaeps}{g_{\vphi e}}

\journal{Physics Letters B}

\begin{document}

\begin{frontmatter}

\title{Electron-seeded ALP production and ALP decay in an oscillating electromagnetic field}

\author{B. King}
\address{Centre for Mathematical Sciences, Plymouth University, Plymouth, PL4 8AA, United Kingdom}

\begin{abstract}
Certain models involving ALPs (axion-like-particles) allow for the coupling of scalars and pseudoscalars to fermions. A derivation of the total rate for production of massive scalars and pseudoscalars by an electron in a monochromatic, circularly-polarised electromagnetic background is presented. In addition, a derivation and the total rate for the decay of massive scalars and pseudoscalars into electron-positron pairs in the same electromagnetic background is given. We conclude by approximating the total yield of ALP production for a typical laser-particle experimental scenario.
\end{abstract}

\end{frontmatter}

\section{Introduction}
The spontaneous breaking of global symmetries in beyond-the-standard-model theories can give rise to scalar or pseudoscalar particles which are commonly referred to as ALPs (axion-like-particles). The original axion is a pseudoscalar that arises when the Peccei-Quinn \cite{peccei77} symmetry is broken at a very high energy scale to give rise to CP violation in QCD, thereby posing a potential solution to the so-called strong-CP problem. Whereas axions that solve the strong-CP problem are bound to this very high energy scale, ALPs are independent and therefore less constrained. The ALP can be hadronic, such as in the original KSVZ axion \cite{kim79,shifman80}, which predicts a new heavy quark and a scalar meson. However, they may also be non-hadronic, such as the DFSZ  axion \cite{zhitnitsky80,dine81} which can couple at the tree level to leptons. Various experimental searches have been and are being conducted to detect ALPs and place increasingly stringent bounds on ALP models. Examples include helioscopes such as CAST \cite{cast17} that use magnetic fields to regenerate axions emitted from the sun, LSW (light-shining-through-the-wall) experiments such as the ALPS experiment \cite{alps10} that use an optical cavity to create ALPs and a regeneration region using inhomogeneous magnetic fields, as well as beam dump KEK \cite{KEK86}, Orsay \cite{orsay89}, E774 \cite{bross91} and fixed-target experiments APEX \cite{apex11} and NA62 \cite{doebrich16} (some reviews of axion searches can be found in \cite{kim10b,andreas12,raffelt13,irastorza18}).

The diphoton-ALP coupling in several ALP models suggests that one may also consider employing sources of large numbers of photons, such as intense laser pulses, for the generation part of a LSW experiment \cite{mendonca07,doebrich10,villalbachavez13,villalbachavez17}. Alternatively, some models couple ALPs directly to electrons \cite{borisov96} which allows for direct production of ALP in collisions of electron beams with intense laser pulses. It is this mechanism that we consider in the current letter.

The QED counterpart of ALP production by an electron in an external EM (electromagnetic) field is Compton scattering. When the EM background is sufficiently intense that the coupling between charge and gauge field can no longer be considered perturbative and arbitrary numbers of interactions must be taken into account, the process corresponds to NLC (Nonlinear Compton Scattering). First studied over sixty years ago \cite{nikishov64,sengupta52,kibble64,harvey09,mackenroth10}, the prediction of an increased \emph{effective} electron mass due to the charge-field coupling, was recently confirmed in experiment \cite{khrennikov15,sakai15,yan17}. The QED counterpart of ALP decay to an electron-positron pair in an EM background is photon-seeded pair-creation \cite{nikishov64,reiss62,nousch12} (photoproduction of a scalar pair in a circularly-polarised monochromatic background has also been considered \cite{villalbachavez12}). The combination of ALP creation and subsequent decay is analogous to the trident process in QED. Although being measured twenty years ago in the landmark E144 experiment at SLAC \cite{burke97}, the trident process in a strong EM background presents numerous theoretical challenges and has recently been the subject of increased interest \cite{hu10,ilderton11,king13b,torgrimsson17,king18a}. Using massive scalars rather than virtual photons offers a simpler system to understand the main issues such as the relative importance of one-step and two-step processes. Reviews of strong-field QED can be found in \cite{ritus85,dipiazza12,narozhny15,king15a}.
\newline

In the current letter, four processes are considered in a monochromatic electromagnetic background: i) massive scalar production by an electron; ii) the decay of a massive scalar to an electron-positron pair and iii) and iv) the same two processes for a pseudoscalar. We present only an outline of the derivation, highlighting steps important to the massive scalar case, as the derivation of equivalent QED processes already exist in the literature \cite{landau4}. Following derivation of the scalar cases, we state the pseudoscalar result.

\begin{figure}[h!!] 
\centering
\begin{subfigure}[b]{2.2cm}
\includegraphics[draft=false,width=\textwidth]{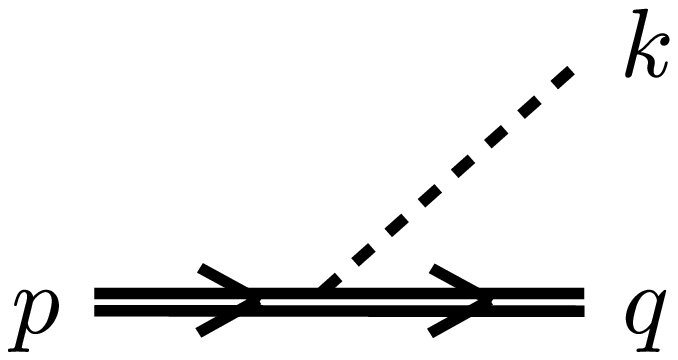}
\caption{Electron-seeded ALP production in an EM background.}
\end{subfigure}%
\hspace{1.5cm} \begin{subfigure}[b]{2.6cm}
\includegraphics[draft=false,width=\textwidth]{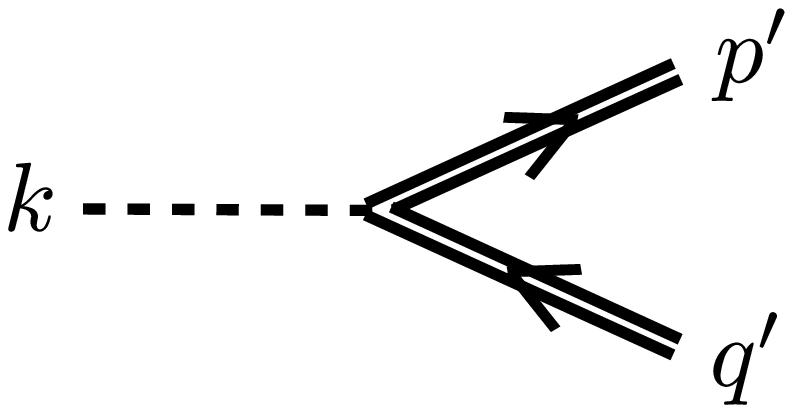} 
\caption{ALP decay to an electron-positron pair in an EM background.}
\end{subfigure}
 \caption{Feynman diagrams for the processes studied. Double lines imply external-field wavefunctions.} \label{fig:diags}
\end{figure}

\section{ALP production by an electron}

The scattering matrix element for the massive scalar production depicted by the Feynman diagram in \figrefa{fig:diags} is:
\bea
 \Sfi = i\gae \int d^{4}x ~\phi_{k}\bar{\psi}_{q} \psi_{p}.
\eea
The coupling of the fermions to the EM background to all orders is incorporated using the Furry picture, in which the S-matrix is expanded as a perturbation around solutions to the Dirac equation in a given EM background (so-called ``dressed states''). The Volkov wavefunction:
\bea
\psi_{p} &=& \left[1+ \frac{\slashed{\vkap}\slashed{a}}{2\vkap\cdot p}\right] \frac{u_{p}}{\sqrt{2p^{0}V}}\e^{-ip\cdot x -iU_{p}(\vphi)}\nn \\
U_{p}(\vphi) &=& \int^{\vphi} \left(\frac{p\cdot a(\phi)}{\vkap \cdot p} - \frac{a^{2}(\phi)}{2\,\vkap\cdot p}\right) d\phi
\eea
describes an electron of charge $e$, momentum $p$ satisfying the on-shell condition $p^{2}=m^{2}$, with free-electron spinor $u_{p}$, propagating in a plane wave background with scaled gauge potential $a(\vphi) = eA(\vphi)$ and phase $\vphi = \vkap \cdot x$, with wavevector $\vkap$ such that $\vkap \cdot a = \vkap\cdot \vkap = 0$. We choose the background to be circularly-polarised and monochromatic:
\bea
a(\vphi) = m\xi \left(\eps\,\cos\vphi +  \epst\,\sin\vphi \right),
\eea
for $\eps\cdot \eps = \epst\cdot\epst = -1$ and $\eps\cdot\epst = 0$, and $\xi$ is the classical nonlinearity parameter \cite{ilderton09} quantifying the strength of the EM background field. We choose the \emph{lab frame} in which $\vkap = \vkap^{0}(1,0,0,1)$, $\eps = (0,1,0,0)$ and $\epst =(0,0,1,0)$. The massive scalar is described by the plane-wave state:
\bea
\phi = \frac{1}{\sqrt{2k^{0}V}}\,\e^{i k\cdot x},
\eea
with momentum $k$ satisfying the on-shell condition $k^{2} = m_{\phi}^{2}$. The form of the scattering matrix element is then:
\bea
 \Sfi \!&=&\! \frac{i\gae}{\sqrt{2^{3}p^{0}q^{0}k^{0}V^{3}}}\! \int\!\! d^{4}x ~\e^{-ix\cdot(p-q-k-\beta\vkap) - i \alpha_{s}\sin\vphi + i \alpha_{c} \cos\vphi} \nn \\
 && \qquad\qquad\quad \times \bar{u}_{q} \left[1+ \frac{\slashed{a}\slashed{\vkap}}{2\vkap\cdot q}\right]  \left[1+ \frac{\slashed{\vkap}\slashed{a}}{2\vkap\cdot p}\right] u_{p},
\eea
where 
\[
\alpha_{s} = m\xi \left(\frac{p\cdot \eps}{\vkap\cdot p}-\frac{q\cdot \eps}{\vkap\cdot q}\right); \quad \alpha_{c} = m\xi \left(\frac{p\cdot \epst}{\vkap\cdot p}-\frac{q\cdot \epst}{\vkap\cdot q}\right)
\]
\[ \beta = -\frac{(m\xi)^{2}}{2}\left(\frac{1}{\vkap \cdot p} - \frac{1}{\vkap \cdot q}\right). \nn \\
\]
It is advisable to deal with the integral after mod-squaring and taking the trace to simplify evaluation of the $x^{-} = \vphi/\vkap^{0}$ integral ($x^{-} = x^{0}-x^{3}$ is the lightfront co-ordinate comoving with the background field). One then arrives at
\bea
 \frac{1}{2}\sum_{\trm{spin}}\tr |\Sfi|^{2} &=& \frac{\gae^{2}}{V^{3}} \int \frac{d^{4}x\, d^{4}x'}{2^{3}p^{0}q^{0}k^{0}} ~\e^{-i(x-x')\cdot(p-q-k-\beta\vkap)} \nn\\
 && \qquad \times \sum_{s,s'}\e^{- i (s-s') \vphi } \,T_{s,s'}
\eea
if the Jacobi-Anger expansion \cite{watson22} is used:
\bea
\e^{-i z \sin (\vphi-\vphi_{0})} &=& \sum_{s=-\infty}^{\infty} \J_{s}(z) \,\e^{-is(\vphi-\vphi_{0})},
\eea
where $\J_{s}(z)$ is the Bessel function of the first kind and in the current calculation, we have:
\[
z^{2} = \alpha_{s}^{2} + \alpha_{c}^{2}; \qquad \cos\vphi_{0} = \frac{\alpha_{s}}{z}; \qquad \sin\vphi_{0} = \frac{\alpha_{c}}{z}
\]
and $T_{s,s'}$ is the result of the trace, which we give later in a more simplified form. Performing the spatial integrals in $x$ and $x'$, one has to deal with the combination:
\[
\delta^{(4)}\left(p-q-k-(\beta-s)\vkap\right)\,\delta^{(4)}\left(p-q-k-(\beta-s')\vkap\right).
\]
This can be simplified into:
\[
\delta^{(4)}\left(p-q-k-(\beta-s)\vkap\right)\,\frac{V}{(2\pi)^{3}}\frac{p^{0}}{\vkap\cdot p} \delta(s-s'),
\]
by considering the combination
\[
\frac{\delta^{(4)}\left((s-s')\vkap\right)}{\delta(s-s')}\Bigg|_{s=s'} = \frac{V}{(2\pi)^{3}}\frac{dt}{d\vphi} = \frac{V}{(2\pi)^{3}}\frac{dt}{d\tau} \frac{d\tau}{d\vphi},
\]
(where $\tau$ is the proper time), and using the result that for an electron in a plane wave EM background, $\vkap \cdot p = m\, d\vphi/d\tau$ is a constant \cite{ilderton09}. Defining the probability for massive scalar production by an electron as $\tsf{P}^{e\to\phi}$, through:
\bea
\tsf{P}^{e\to\phi} = V^{2}\int \frac{d^{3}k}{(2\pi)^{3}}\,\frac{d^{3}q}{(2\pi)^{3}} \frac{1}{2}\sum_{\trm{spin}}\tr |\Sfi|^{2},
\eea
one has the intermediate step:
\bea
\tsf{P}^{e\to\phi} &=& \frac{\gae^{2}}{4\pi \vkap\cdot p} \int \frac{d^{3}k}{2k^{0}}\,\frac{d^{3}q}{2q^{0}}\sum_{s,s'} \delta(s-s') \nn \\ && \qquad \times~
\delta^{(4)}\left(p-q-k-(\beta-s)\vkap\right) T_{s,s'}. \label{eqn:int1}
\eea
Here we notice the clear appearance of global momentum conservation:
\[
\tilde{p} + s\vkap = \tilde{q} + k,
\]
where we define the electron \emph{quasimomentum}, $\protect{\tilde{p} = p - (a^{2}/p\cdot \vkap) \vkap}$, and the relation to the \emph{effective} mass $m_{\ast}$ via $\tilde{p}^{2} = m_{\ast}^{2}$, giving $m_{\ast}^{2} = m^{2}(1+\xi^{2})$ for a circularly-polarised background \cite{lavelle15,lavelle17}. Moreover the integer $s$ is suggestive as the number of photons of frequency $\vkap^{0}$ absorbed from the EM background. Let us now deal with the final delta-function by recognising:
\[
\sum_{s,s'} \delta(s-s') = \int \frac{d\vphi}{2\pi} \sum_{s},
\]
and since the phase integral is divergent, let us define the rate per phase $\tsf{R}^{e\to\phi} = \tsf{P}^{e\to\phi}/\int d\vphi$. In addition, let us write this rate as a sum over the rate for each harmonic $s$:
\[
\tsf{R}^{e\to\phi} = \sum_{s=s_{0}^{\phi}}^{\infty}\tsf{R}^{e\to\phi}_{s},
\]
where $s_{0}^{\phi}$ is some threshold integer number of photons (which we will later define in terms of the integration variable), above which a scalar with mass $\sqrt{k^{2}}$ can be produced by the electron. If the $q$ integral is performed in \eqnref{eqn:int1}, we have:
\bea
\tsf{R}^{e\to\phi}_{s} &=& \gae^{2}\! \int \! d^{4}k~\frac{\theta(k^{0})\theta(q^{0})}{8\pi^{2} \vkap\cdot p}\delta(k^{2}-m_{\phi}^{2})\delta(q^{2}-m^{2})T_{s,s}\nn \\\label{eqn:int2}
\eea
\bea
T_{s,s} &=& \left(4m^{2} - m_{\phi}^{2} \right)\J_{s}^{2}(z) \nn \\
&& + \frac{(m\xi)^{2}(k\cdot\vkap)^{2}}{2\,p\cdot\vkap\,q\cdot \vkap} \left( \J_{s+1}^{2}(z) + \J_{s-1}^{2}(z) - 2\J_{s}^{2}(z)\right). \nn 
\eea
Suppose the $k$ integral is written as $d^{4}k = dk^{+}dk^{-}d|k^{\perp}|^{2}d\psi/4$, where $k^{\pm} = k^{0} \pm k^{3}$ are lightfront co-ordinates and $\psi$ is the polar angle in the $k^{\perp} = (k^{1}, k^{2})$ plane. Then by using the remaining delta-functions to integrate in $|k^{\perp}|^{2}$ and $k^{+}$, it can be shown the integrand is independent of the angle $\psi$. We then have:
\bea
\tsf{R}^{e\to\phi}_{s} &=& \frac{\gae^{2}}{16\pi} \int \frac{d (\vkap \cdot k)}{(\vkap \cdot p)^2}\,\theta(\vkap\cdot k)\theta(\vkap\cdot p-\vkap\cdot k)\,T_{s,s}.\nn \\
\eea
In order to make a comparison with literature results on the QED process of NLC, we rewrite this integration as:
\bea
\tsf{R}^{e\to\phi}_{s} &=& \frac{\gae^{2}}{16\pi \eta_{p}}\int_{u_{s}^{-}}^{u_{s}^{+}} \frac{du}{(1+u)^{2}}\left\{\left(4 - \delta^{2} \right)\J_{s}^{2}(z_{s}^{\phi}) \right.\nn \\
&&  \left. \quad+ \frac{u^{2}}{2(1+u)} \left[ \J_{s+1}^{2}(z_{s}^{\phi}) + \J_{s-1}^{2}(z_{s}^{\phi}) - 2\J_{s}^{2}(z_{s}^{\phi})\right]	\right\},\nn \\ \label{eqn:Rphis}
\eea
where we define the energy parameter $\eta_{p}  = \vkap \cdot p /m^{2}$, and the ALP mass parameter $\delta = m_{\phi}/m$, and where:
\bea
\left(z^{\phi}_{s}\right)^{2} = \left(\frac{2s\xi}{\sqrt{1+\xi^{2}}}\right)^{2} \frac{u}{u_{s}}\left(1-\frac{u}{u_{s}}\right) - \frac{\delta^{2}\xi^{2}(1+u)}{\eta_{p}^{2}}, \label{eqn:zz}
\eea
and $ u= \eta_{k}/\eta_{q}$ with $\eta_{q} = \eta_{p}-\eta_{k}$ and $u_{s} = 2s \eta_{p}^{\ast}$ with $\eta_{p}^{\ast} = \vkap \cdot p / m_{\ast}^{2}$. Since $z^{2} > 0$ and $z \in \mathbb{R}$, a condition is placed upon the range of integration of the variable $u>0$, namely that the integration bounds $u_{s}^{\pm}$ are given by:
\bea
u_{s}^{\pm} = \frac{2s\eta_{p}-\delta^{2}}{2(1+\xi^{2})} \left[1 \pm \sqrt{1-\frac{4(1+\xi^{2})\delta^{2}}{(2s\eta_{p}-\delta^{2})^{2}}}\right]
\eea
where we choose the range of parameters in which $u_{s}^{+} \geq u_{s}^{-} \geq 0$. We note that in the zero-mass limit, $\delta \to 0$, $u^{+}_{s} \to u_{s}$ and $u^{-}_{s} \to 0$ and $z_{s}^{\phi}$ tends to the standard argument for NLC of a massless photon in a circularly-polarised monochromatic background \cite{landau4}. In the zero-mass limit, the form of the integrand is slightly different to the standard NLC case. First, the coefficient of the first $\J_{s}^{2}$ term has a different sign and the factor $2$ coefficient before the bracket of three squared Bessel functions is missing. Both of these originate from the different trace of a the electron-scalar interaction compared with the electron-photon interaction in QED. Second, the coefficient of the entire integral is a factor $1/4$ smaller than the analogous QED result \cite{bamber99}. This is the well-known factor that originates from the missing polarisation sum in a scalar analogue of a photon interaction. (Recent calculations of NLC in a monochromatic background in scalar QED demonstrate the same $1/4$ pre-factor \cite{king16c,raicher16}.)

\begin{figure}[h!!] 
\centering
\begin{subfigure}[t]{4.0cm}
\includegraphics[draft=false,width=\textwidth]{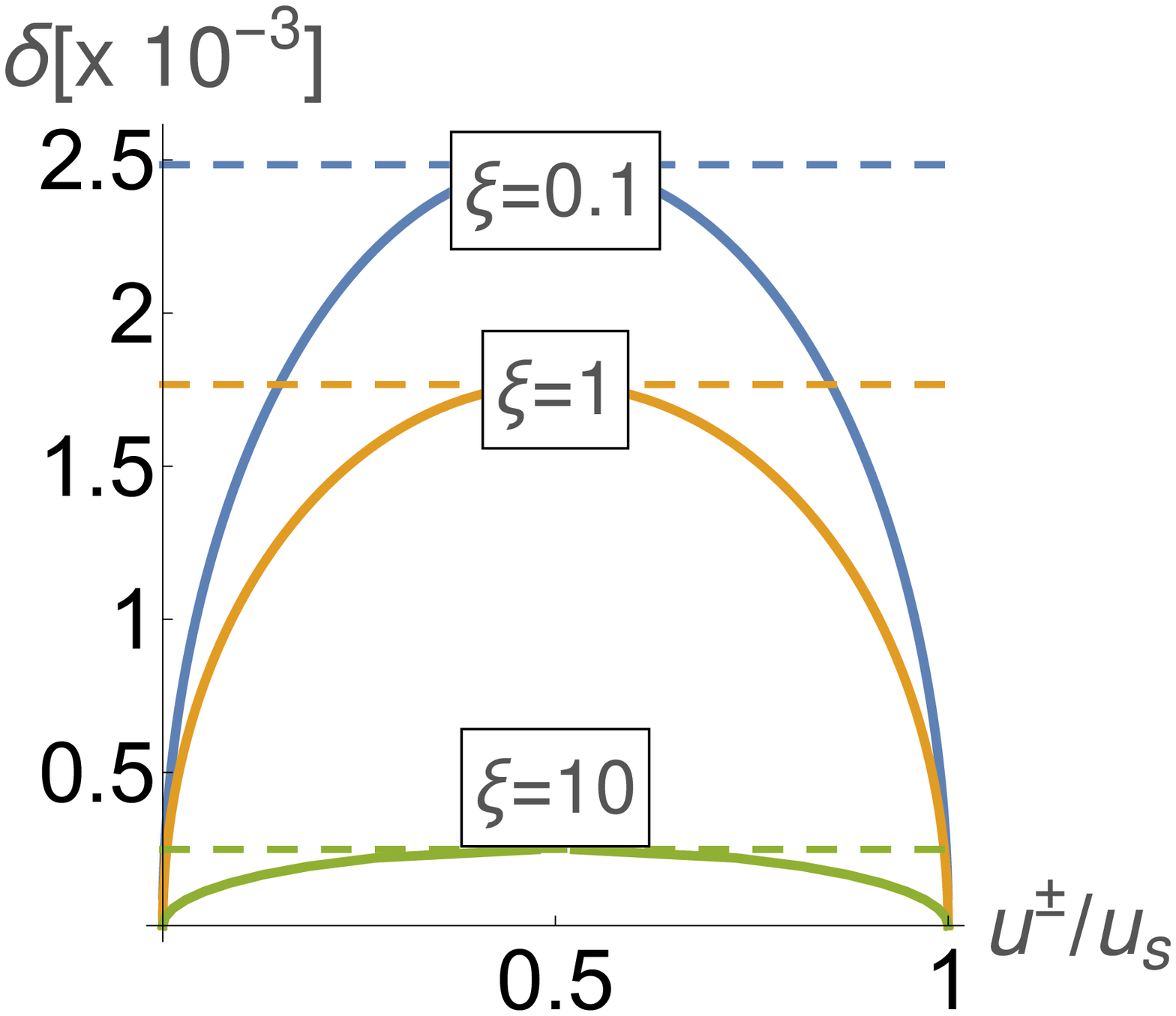}
\caption{How the harmonic range for the fundamental harmonic ($s=1$) shrinks as the scalar mass increases.}
\end{subfigure}
\hspace{0.5cm}
\begin{subfigure}[t]{4.0cm}
\includegraphics[draft=false,width=\textwidth]{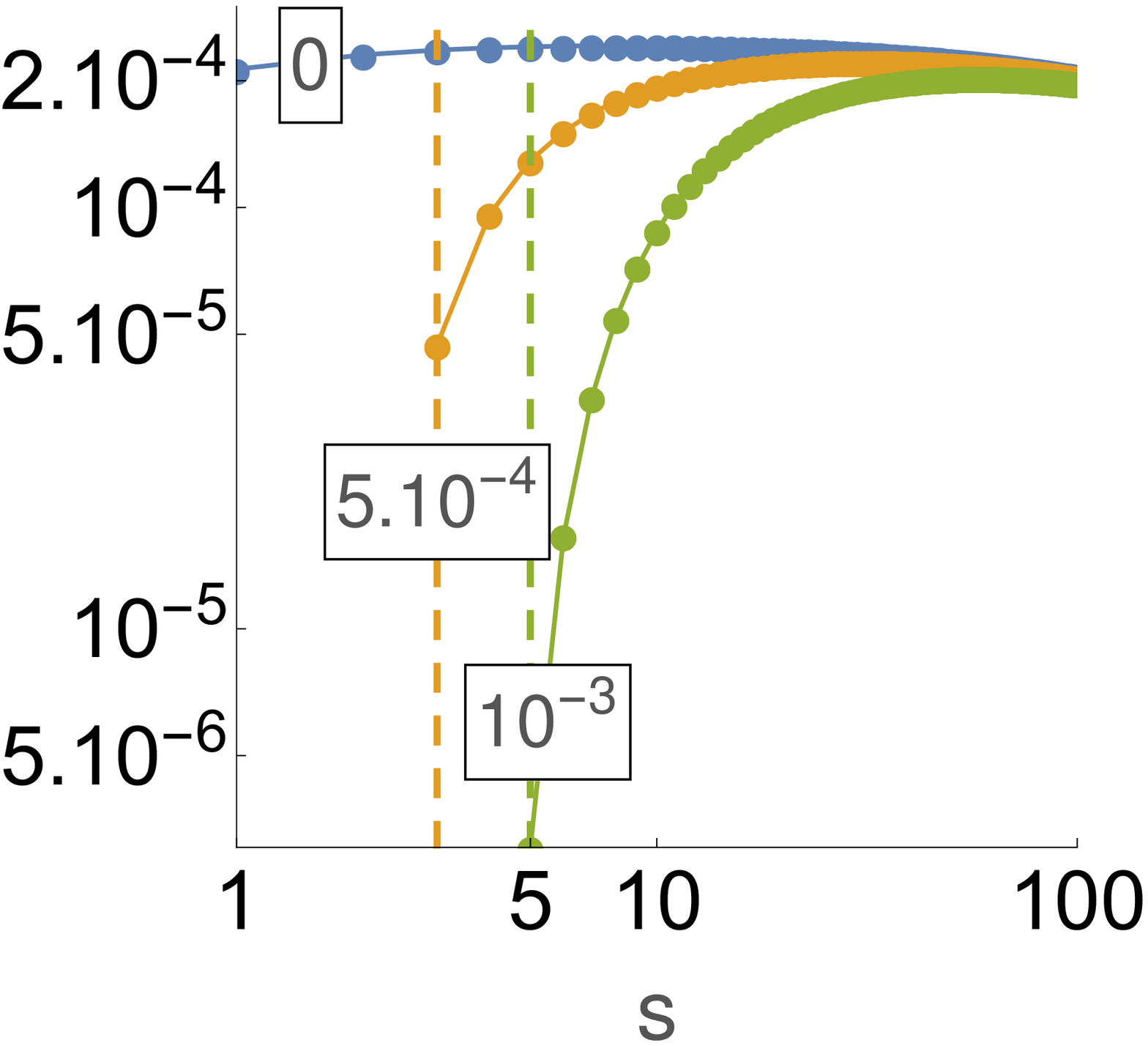} 
\caption{Suppression of lower harmonics in a plot of $R_{s}^{e\to\phi}$ against harmonic order $s$ for $\xi=10$, when the scalar mass (indicated by curve labels) is increased.}
\end{subfigure}
 \caption{The effect of the scalar mass on scalar production by an electron for $\eta_{p}=0.0025$ ($g_{\phi e} = 1$).} \label{fig:anlc}
\end{figure}

Just as in NLC, there exists an interval in $u$ for each harmonic, the so-called ``harmonic range''. The difference here is that when the scalar mass is increased, this harmonic range is reduced on both sides, as displayed in \figrefa{fig:anlc}. The heaviest scalar that can be produced at a given harmonic has a mass parameter:
\bea
\delta = 2\left(1+ s\eta_{p} + \xi^{2} - \sqrt{(1+\xi^{2})(1+2s\eta_{p} + \xi^{2})}\right),
\eea  
indicated by the horizontal dashed lines in \figrefa{fig:anlc} (this can be derived by finding the value of $\delta$ such that $u_{s}^{+}=u_{s}^{-}$). Furthermore, in \figrefa{fig:anlc} it is shown how this suppression of the range affects larger values of $\xi$ more than smaller ones (this trend is continued for higher harmonics). 

Another effect of a non-zero scalar mass is the appearance of a threshold number of photons that must be taken from the background field before the scalar can be scattered. By squaring the centre-of-mass energy, we find that:
\bea
2s\eta_{p} = \delta^{2}\frac{\eta_{p}}{\eta_{k}} + \frac{m_{\ast}^{2}}{m^{2}}\frac{\eta_{k}}{\eta_{q}} + \frac{\left(q^{\perp}\eta_{k} - k^{\perp}\eta_{q}\right)^{2}}{\vkap\cdot q~\vkap\cdot k}. \label{eqn:s0deriv}
\eea
Therefore the threshold number of photons required to create a scalar with mass $\delta$ is $s^{\phi}_{0}$ where
\bea
s^{\phi}_{0} = \left\lceil\frac{\delta}{\eta_{p}}\left(\frac{\delta}{2} + \sqrt{1+\xi^{2}}\right)\right\rceil
\eea
($\lceil \cdot \rceil$ denotes the ceiling function), which reduces to  $s^{\phi}_{0}=0$ in the massless scalar limit, analogous to NLC of massless photons in QED.
The effect on the harmonic rate of having a threshold number of photons for the scattering of a massive scalar is evident in \figrefb{fig:anlc}. It is also apparent that even for many harmonics above the threshold, the scalar production rate is still suppressed compared to the zero-mass case. Also from \figrefb{fig:anlc}, it is clear that the higher the scalar mass, the larger the suppression for the above-threshold harmonics.
\newline

The scattering matrix element for massive pseudoscalar production can be written \cite{borisov96}:
\bea
 \Sfi = i\gaeps \int d^{4}x ~\varphi_{k}\bar{\psi}_{q} \gamma_{5} \psi_{p},
\eea
where we denote a pseudoscalar by $\vphi$. The derivation of the total rate follows identical lines. If we define the total pseudoscalar rate $\tsf{R}^{e\to\vphi} = \sum_{s>s_{0}^\phi}\tsf{R}^{e\to\vphi}_{s}$, then we find:
\bea
\tsf{R}^{e\to\vphi}_{s} &=& \frac{\gaeps^{2}}{16\pi \eta_{p}}\int_{u_{s}^{-}}^{u_{s}^{+}} \frac{du}{(1+u)^{2}}\left\{- \delta^{2}\J_{s}^{2}(z_{s}^{\phi}) \right.\nn \\
&&  \left. \quad+ \frac{u^{2}}{2(1+u)} \left( \J_{s+1}^{2}(z_{s}^{\phi}) + \J_{s-1}^{2}(z_{s}^{\phi}) - 2\J_{s}^{2}(z_{s}^{\phi})\right)	\right\},\nn \\
\eea
where the difference to the massive scalar case \eqnref{eqn:Rphis} is entirely due to the different sign of the mass term in the trace. As the kinematics are the the same, so is the Bessel argument $z_{s}^{\phi}$, already given through \eqnref{eqn:zz}.
\newline

\subsection{Weak-field limit of ALP production by an electron}
If $\xi\ll 1$, the background plane wave is termed ``weak'' and the coupling to electron and positron states is perturbative. For the analogous QED process of NLC, the fundamental harmonic ($s=1$) of the weak-field limit of a monochromatic background is identical to the Klein-Nishina formula \cite{landau4}. Since $z^{\phi}_{s} \propto \xi$, we can arrive at the weak-field limit using the replacements:
\bea
 \J_{s+1}^{2}(z^{\phi}_{s}) + \J_{s-1}^{2}(z^{\phi}_{s})-2\J_{s}^{2}(z^{\phi}_{s}) &\approx& 1 + O(\xi^{2});\nn \\
  \J_{s}^{2}(z^{\phi}_{s}) &\approx& \frac{(z^{\phi}_{s})^{2}}{4} + O(z^{4}).\label{eqn:Jsapp}
\eea
This then leads to:
\bea
\textsf{R}^{e\to\phi}_{1} &\approx& \frac{\gae^{2}\,\xi^{2}}{8\pi\,\eta_{p}}\int_{u_{s}^{-}}^{u_{s}^{+}} \frac{du}{(1+u)^{2}}\left\{\frac{1}{2}\frac{u^{2}}{1+u} \right. \nn 
\\ &&\left. \quad+ (4-\delta^{2})\left[\frac{u}{u_{1}}\left(1-\frac{u}{u_{1}}\right) - \frac{\delta^{2}(1+u)}{4\eta_{p}^{2}}\right] \right\}.\nn \\
\eea
An expansion of the result to leading order in $\delta^{2}$ gives:
\bea
\tsf{R}^{e\to\phi}_{1} &\approx& \frac{\gae^{2}\,\xi^{2}}{8\pi \eta_{p}}\left\{\frac{1}{(1+u_{1})^{2}}\left[\frac{3}{4}\,u_{1}^{2}+\frac{17\,u_{1}}{2} +  16 + \frac{8}{u_{1}} \right] \right.\nn \\
&& \left. -\left[\frac{1}{2}+\frac{4}{u_{1}} + \frac{8}{u_{1}^2} \right]\ln(1+u_{1})\right. \nn \\
&& \left. + \delta^{2}\left(-\frac{2}{u_{1}} +\left[\frac{1}{\eta_{p}^{2}}+\frac{1}{u_{1}} + \frac{2}{u_{1}^{2}} \right]\ln(1+u_{1})\right)\right\}. \nn \\
\label{eqn:wfanlc}
\eea
The pseudoscalar result is then:
\bea
\tsf{R}^{e\to\vphi}_{1} &\approx& \frac{\gaeps^{2}\,\xi^{2}}{8\pi \eta_{p}}\left\{\frac{1}{(1+u_{1})^{2}}\left[\frac{3}{4}\,u_{1}^{2}+\frac{u_{1}}{2}\right]-\frac{1}{2}\ln(1+u_{1}) \right.\nn \\
&& \left. + \delta^{2}\left(-\frac{2}{u_{1}} +\left[\frac{1}{u_{1}} + \frac{2}{u_{1}^{2}} \right]\ln(1+u_{1})\right)\right\}. \nn \\
\label{eqn:wfpsanlc}
\eea

(In a recent calculation of linear Compton production of a massive scalar and pseudoscalar in an external plane-wave field \cite{king18b}, the long-pulse limit was found to agree with \eqnrefs{eqn:wfanlc}{eqn:wfpsanlc} when the background was circularly-polarised.)

\section{ALP decay to electron-positron pair}
The scattering matrix element for the decay of a massive scalar to an electron-positron pair depicted by the Feynman diagram in \figrefb{fig:diags} is:
\bea
 \Sfi = i\gae \int d^{4}x ~\phi_{k}\bar{\psi}_{p'} \psi_{q'}^{+} ,
\eea
where the outgoing positron Volkov wavefunction is
\bea
\psi_{q'}^{+} &=& \left[1- \frac{\slashed{\vkap}\slashed{a}}{2\vkap\cdot q'}\right] \frac{v_{q'}}{\sqrt{2q'^{0}V}}\e^{iq'\cdot x -iU_{-q'}(\vphi)},
\eea
for free-positron spinor $v_{q'}$. The derivation follows very much the structure of the previous section. Let us define the total rate for a single massive scalar to decay to an electron-positron-pair as:
\[
\tsf{R}^{\phi\to e} = \sum_{s=s_{0}^{e}}^{\infty}\tsf{R}^{\phi\to e}_{s}.
\]
Just as for \eqnref{eqn:s0deriv}, but for the pair-creation kinematics, squaring the centre-of-mass energy leads to a threshold number of photons:
\bea
s^{e}_{0} = \left\lceil \frac{1}{2\eta_{k}}\left(\frac{4m^{2}_{\ast}}{m^{2}} - \delta^{2} \right)\right\rceil.
\eea
In other words, the heavier the scalar, the lower the threshold for pair-creation. We then find:
\bea
\tsf{R}^{\phi\to e}_{s} &=& \frac{\gae^{2}}{16\pi \eta_{k}}\int_{1}^{v_{s}^{+}} \frac{dv}{v\sqrt{v(v-1)}}\left\{\left(-4 + \delta^{2} \right)\J_{s}^{2}(z^{e}_{s}) \right.\nn \\
&&  \left. \quad+ 2v\xi^{2}\, \left[ \J_{s+1}^{2}(z^{e}_{s}) + \J_{s-1}^{2}(z^{e}_{s}) - 2\J_{s}^{2}(z^{e}_{s})\right]	\right\},\nn \\ \label{eqn:Rphies}
\eea
where, to make contact with the QED result \cite{landau4}, we have used the integration variable $v = \eta_{k}^{2}/4\eta_{p'}\eta_{q'}$, and
\bea
\left(z^{e}_{s}\right)^{2} = 8\left(\frac{v\,\xi}{\eta_{k}}\frac{m_{\ast}}{m}\right)^{2}\left(\frac{v_{s}^{+}}{v}-1\right),
\eea
where:
\[
 v_{s}^{+} = v_{s}^{0} + \frac{\delta^{2}}{4(1+\xi^{2})}; \quad v_{s}^{0} = \frac{s\,\eta_{k}}{2}\frac{m^{2}}{m_{\ast}^{2}}.
\]
(We note that the relation between $v$ and the lightfront momentum $p'^{-}$ (which occurs naturally when integrating over outgoing particle momenta of the mod-squared scattering matrix element), is nonlinear and splits the original integral into two identical branches.)
\begin{figure}[h!!] 
\centering
\begin{subfigure}[t]{4.0cm}
\includegraphics[draft=false,width=\textwidth]{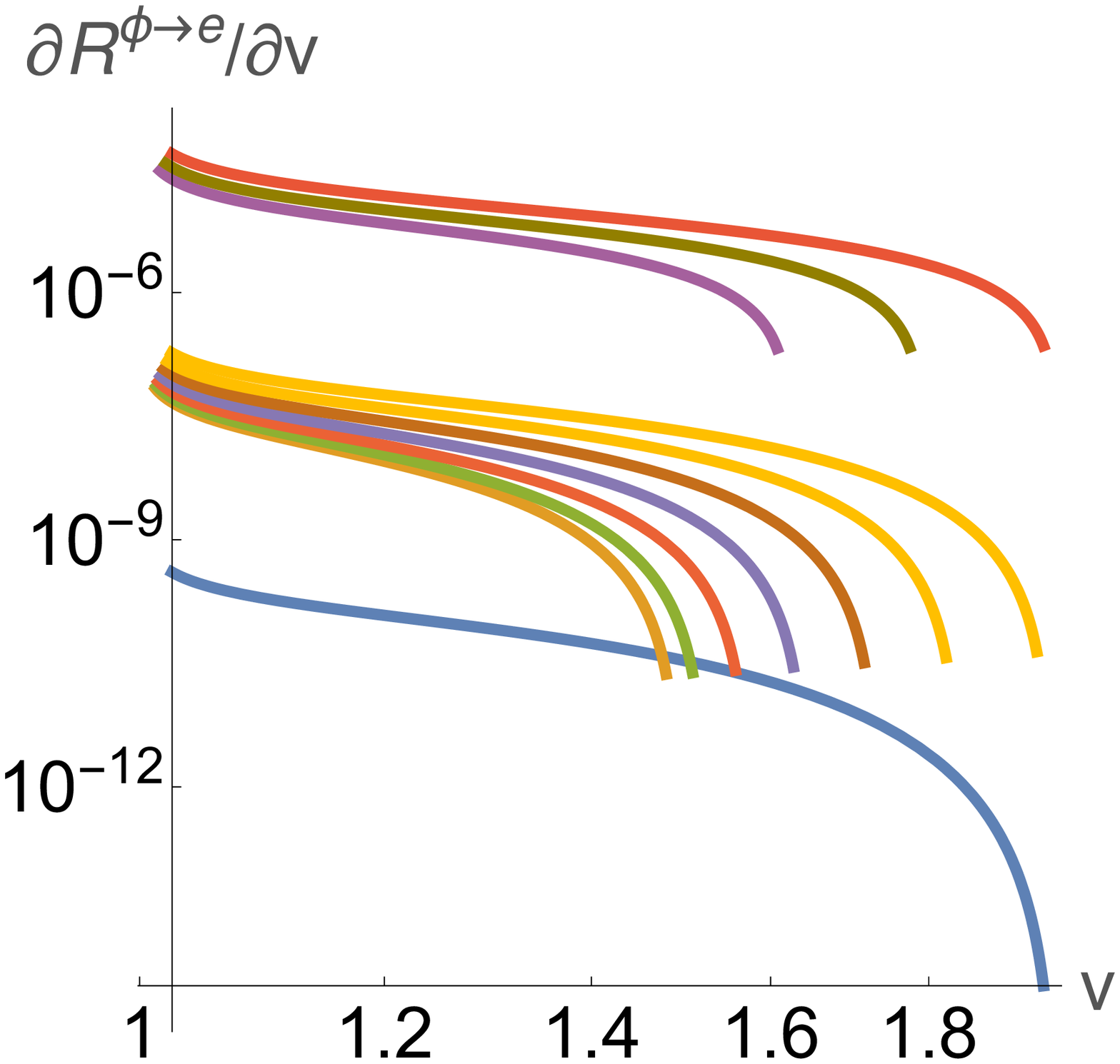} 
\caption{Plot of differential rate of stimulated pair-creation for the harmonic one higher than the threshold $s_{e}^{\phi}$ for $\xi=0.1$, $\eta_{k}=1$ and $\delta$ evenly spaced between $0$ and $2$ ($g_{\phi e} = 1$).}
\end{subfigure}
\hspace{0.5cm}
\begin{subfigure}[t]{4.0cm}
\includegraphics[draft=false,width=\textwidth]{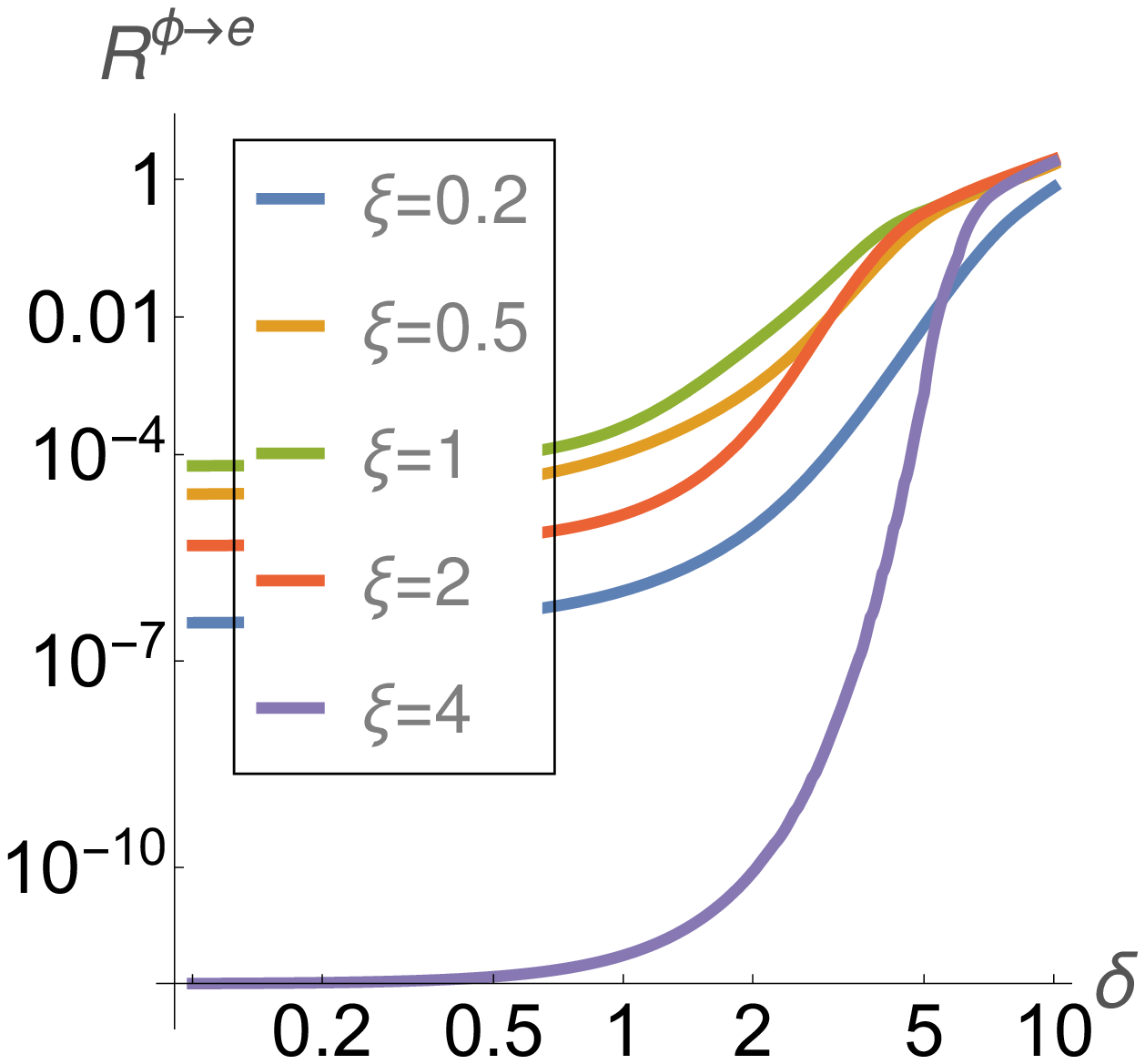}
\caption{How the behaviour of the total pair-creation rate changes when $\xi$ is increased from $\xi<1$ to $\xi>1$ ($\eta_{k}=1$).}
\end{subfigure}
 \caption{The dependency of pair-creation on scalar mass.}\label{fig:pc}
\end{figure}
How the kinematic range and probability change for pair-creation when the scalar mass parameter $\delta$ is increased from $0$ to $2$ in equal increments, is shown in \figrefa{fig:pc}. The lowest curve is for $\delta=0$ and corresponds to a threshold $s_{e}^{\phi}=3$ photons. For $\delta=0.2$, the threshold then drops to $s_{e}^{\phi}=2$ photons, and as $\delta$ is further increased, the kinematic range opens up and the curves become wider, whilst staying at approximately the same amplitude. When $\delta$ reaches $1.6$, the threshold drops to $1$ photon. When $\delta=2$, the process of scalar decay can occur without the background field and as $\delta$ is raised above this, the process moves from being \emph{field-induced} to \emph{field-free} or from \emph{stimulated} to \emph{spontaneous} decay.

The value of $\xi=1$ is particularly important in pair-creation. As can be seen from \eqnref{eqn:Rphies}, for $\xi \ll 1$, the main contribution is given by the first term in the braces, whereas for $\xi \gg 1$, one expects the second combination of Bessel functions to dominate. As $\xi$ is increased from $0.2$ to $1$, the dependence of the total rate on the scalar mass is similar, and the total rate increases with $\xi$. However, for $\xi>1$, the dependence of the total rate on $\delta$ becomes more sensitive and for $\delta <2$ the rate is suppressed much more than for the $\xi<1$ cases, as shown in \figrefb{fig:pc}.
\newline

For the case of pseudoscalar decay to an electron-positron pair in a monochromatic background, the scattering matrix element is given by:
\bea
 \Sfi = i\gaeps \int d^{4}x ~\phi_{k}\bar{\psi}_{p'} \gamma_{5}\psi_{q'}^{+}.
\eea
As for ALP production, in the case of ALP decay, it is only the mass term that changes sign between the scalar and pseudoscalar cases. Therefore, we skip straight to the result for the total pseudoscalar rate $\tsf{R}^{\vphi\to e} = \sum_{s>s_{0}^{e}}\tsf{R}^{\vphi\to e}_{s}$ where:
\bea
\tsf{R}^{\vphi\to e}_{s} &=& \frac{\gaeps^{2}}{16\pi \eta_{k}}\int_{1}^{v_{s}^{+}} \frac{dv}{v\sqrt{v(v-1)}}\left\{\delta^{2} \J_{s}^{2}(z^{e}_{s}) \right.\nn \\
&&  \left. \quad+ 2v\xi^{2}\, \left[ \J_{s+1}^{2}(z^{e}_{s}) + \J_{s-1}^{2}(z^{e}_{s}) - 2\J_{s}^{2}(z^{e}_{s})\right]	\right\}.\nn \\ \label{eqn:Rvphies}
\eea
For light pseudoscalars with $\delta \ll 1$, we see from \eqnref{eqn:Rvphies} a large suppression in the rate. Therefore light pseudoscalars are much more stable than light scalars when propagating through weak EM backgrounds.

\subsection{Weak-field limit of ALP decay to electron-positron pair}
Using the same expansion as \eqnref{eqn:Jsapp} when $\xi \ll 1$, we find the weak-field limit of the ALP decay in a monochromatic background to be given by:
\bea
\tsf{R}^{\phi\to e}_{s} &\approx& \frac{\gae^{2}\,\xi^{2}}{8\pi\,\eta_{k}^{3}}\left\{-\telta^{2}\sqrt{t_{s}(t_{s}+1)}  \right.\nn \\
&& \left. + \left[2\eta_{k}^{2} + \telta^{2}(2t_{s}-1)\right]\sinh^{-1}\!\sqrt{t_{s}}\right\}
\eea

where we have defined the shorthand: $\telta^{2} = \delta^{2}-4$ and the Mandelstam-like variable $t_{s} = -1+(k+s\vkap)^{2}/4m^{2}$, equal to the relative difference of the centre-of-mass energy squared to the pair rest energy squared.

\section{Discussion}
Although the diagram for ALP production by an electron in an EM plane wave is very similar to nonlinear Compton scattering, when a massive scalar is emitted, the kinematics are more akin to pair-creation in an EM plane wave. In particular, there appears a threshold number of external-field photons that depends on the scalar's mass, the frequency of the background and the energy of the electron.
\newline

Pair creation in a plane wave EM background by a massive scalar differs from pair-creation from a photon, in that the mass of the scalar lowers the threshold number of photons required for the process to proceed. As the scalar mass is increased, the process changes from being a stimulated to a spontaneous process.
\newline

\begin{figure}[h!!] 
\centering
\begin{subfigure}[t]{4.0cm}
\includegraphics[draft=false,width=4cm]{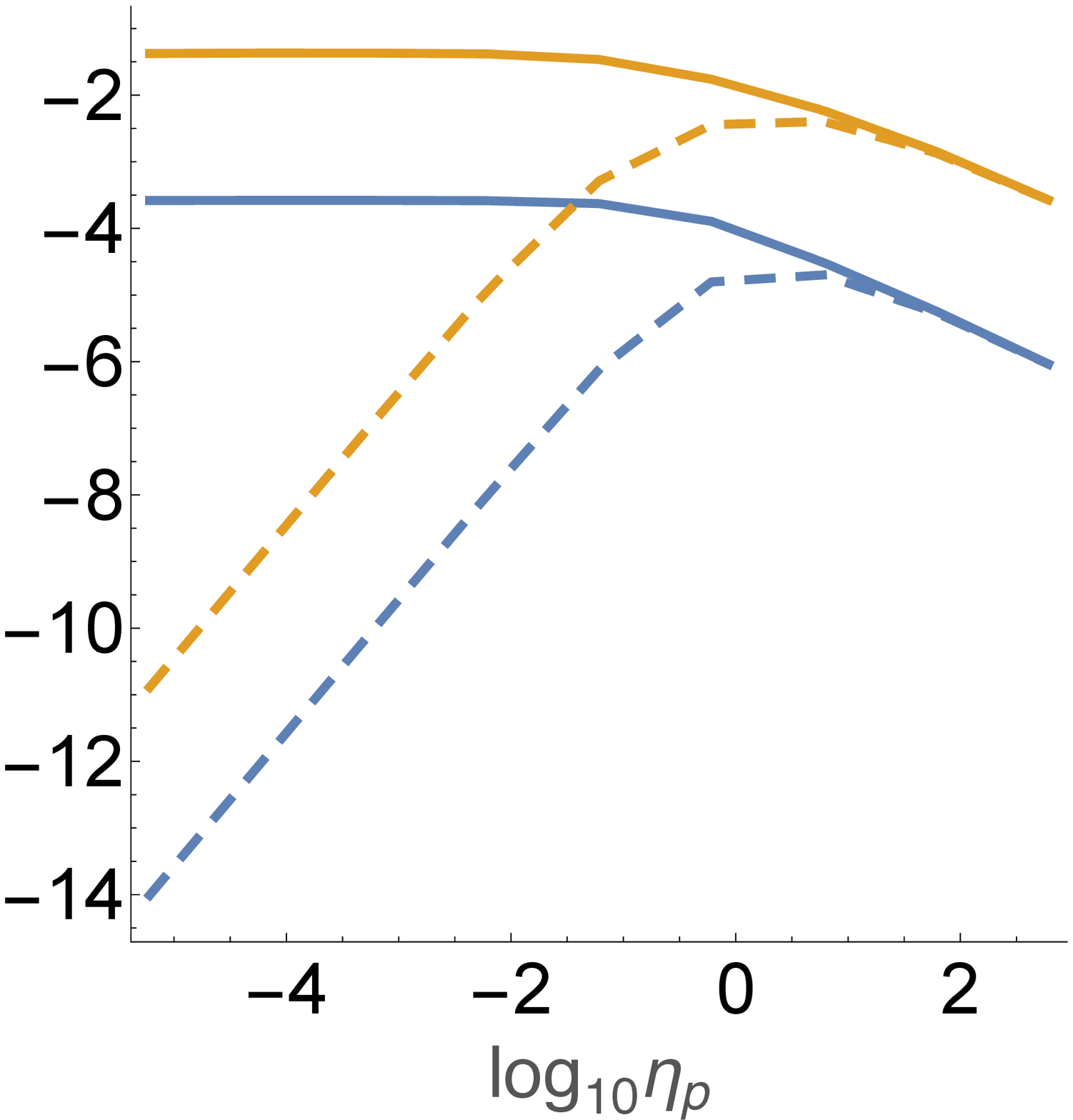} 
 \caption{A log-log plot of how the pseudoscalar rate (dashed lines) and scalar rates (solid lines) depend on $\eta_{p}$ for $\xi=2$ (upper pair of coalescing lines) and $\xi=0.1$ (lower pair of coalescing lines).}\label{fig:pssp}
 \end{subfigure}
 \hspace{0.5cm}
 \begin{subfigure}[t]{4.0cm}
\includegraphics[draft=false,width=4cm]{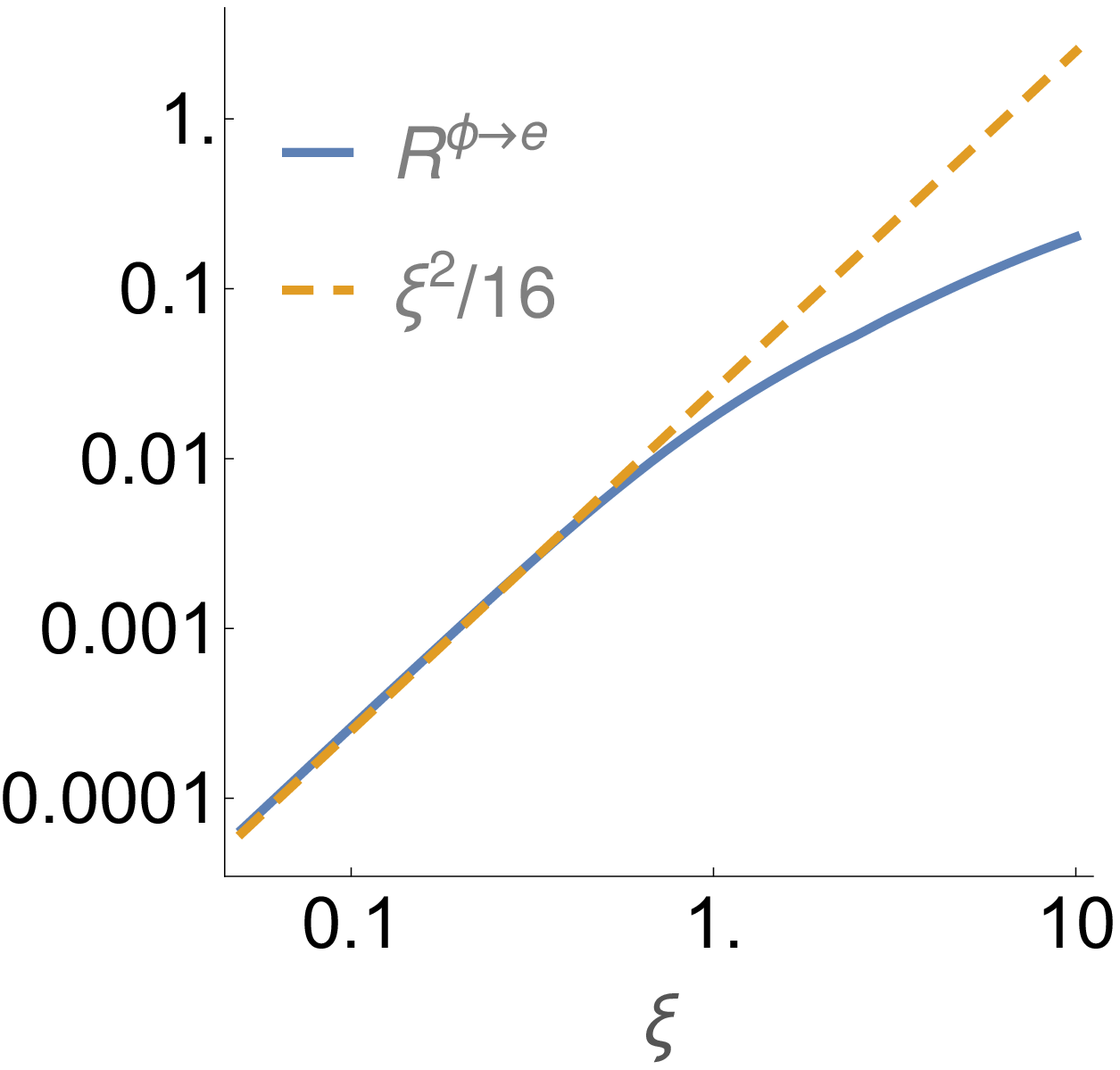} 
 \caption{The dependency of pair-creation on scalar mass and the fit $R^{\phi\to e} =0.025\xi^{2}$ (dashed) for $\gae=1$, and $\eta_{p} = 6 \cdot 10^{-6}$, $m_{\phi}=1\trm{meV}$ (making $\delta = 2\cdot 10^{-9}$).} 
 \end{subfigure}
 \caption{Dependency of total scalar rate on $\xi$ and $\eta_{p}$.}\label{fig:NLCtot}
\end{figure}

We do not perform here an analysis of the ALP production and regeneration rates expected in experiment, however we give some orders of magnitude for the production mechanism. The dependency of the total rate for massless scalar production by an electron is shown in \figref{fig:NLCtot} for $\gae=1$, and $\eta_{p} = 6 \cdot 10^{-6}$ (equivalent to an electron at rest in a monochromatic background of frequency $\vkap^{0} = 1.55\,\trm{eV}$  -- the frequency of a $800\,\trm{nm}$ laser beam). It can be seen in \figrefa{fig:NLCtot} that as $\eta_{p} \to \infty$, the pseudoscalar and scalar rates tend to the same values. A similar behaviour was found in the monochromatic limit of a direct calculation of these processes in a background pulse \cite{king18b}.

To estimate the total number of scalars produced, we introduce for $\eta_{p} = 6\cdot 10^{-6}$, the weak-field fit: $R^{\phi\to e} \approx 0.025\xi^{2}$, the accuracy of which is plotted in \figrefb{fig:NLCtot}. (This scaling with $\xi$ is clear from the weak-field result in \eqnref{eqn:wfanlc}.) Although we have picked a specific value of $\eta_{p}$, we see from \figrefa{fig:NLCtot}, that for all $\eta_{p}\lesssim 10^{-2}$, the rate of scalar production is not very sensitive to the value of $\eta_{p}$. Therefore, for $\xi \ll 1$, we can estimate the yield per collision as $N_{\phi} = 0.025 N_{e} \xi^{2} g_{\phi e}^{2} \Phi$ or:
\[
N_{\phi} \approx 4 \times 10^{8}\,g_{\phi e}^{2}\,\left(\frac{N_{e}}{10^{8}}\right)\left(\frac{I}{10^{19}\,\trm{Wcm}^{-2}}\right)\left(\frac{1.55\,\trm{eV}}{\vkap^{0}}\right)\left(\frac{\tau}{100\,\trm{fs}}\right)
\]
where $\Phi = \vkap^{0}\tau$ is the laser pulse phase length and $\tau$ is the pulse duration, and where $N_{e}$ is the number of electron seeds. If one takes a \emph{lab-based} limit on $\gae$ from underground detectors of the order $\sim O(10^{-11})$ \cite{irastorza18} (astrophysical bounds are of the order of $\sim O(10^{-13})$ for axions in the meV range \cite{redondo13}), and an optical long pulse duration of $\Phi \sim O(10^{3})$ or X-ray long pulse duration of $\Phi \sim O(10^{6})$, then to achieve $N_{\phi}\gg1$, it is beneficial to use a large number of electron seeds. Laser-wakefield-accelerated electron beams containing of the order of $10^{9}-10^{10}$ electrons have been demonstrated in the lab \cite{mcguffey10,esarey09}. Moreover, as we see from \figref{fig:NLCtot}, the initial momentum of the electrons is not crucial. The limiting factor is more the volume of the laser pulse for which $\xi$ is sufficiently high, as well as the laser's repetition rate. The BELLA laser facility supplies $1\,\trm{PW}$ at a repetition rate of $1\,\trm{Hz}$ and has demonstrated a peak laser intensity of over $10^{19}\,\trm{Wcm}^{-2}$ \cite{nakamura17}. With the development of modern high-intensity laser systems, these values are set to become even more favourable.

  Therefore, the experimental set-up of colliding a high rep-rate intense laser pulse with an electron gas may not immediately provide as stringent limits as astrophysical bounds, however, by varying the electron-beam parameters, a large range of scalar masses can be probed. Moreover, such an experiment would be unique, in that it would provide the first fully lab-based, model-independent probe of $g_{\phi e}$.

\section*{Acknowledgments}
The author acknowledges discussions with B. M. Dillon and funding from Grant No. EP/P005217/1.

\section*{References}

\bibliography{current}

\providecommand{\noopsort}[1]{}
\begin{thebibliography}{10}
\expandafter\ifx\csname url\endcsname\relax
  \def\url#1{\texttt{#1}}\fi
\expandafter\ifx\csname urlprefix\endcsname\relax\def\urlprefix{URL }\fi
\expandafter\ifx\csname href\endcsname\relax
  \def\href#1#2{#2} \def\path#1{#1}\fi

\bibitem{peccei77}
R.~D. Peccei, H.~R. Quinn,
  \href{https://link.aps.org/doi/10.1103/PhysRevLett.38.1440}{$\mathrm{CP}$},
  Phys. Rev. Lett. 38 (1977) 1440--1443.
\newblock \href {http://dx.doi.org/10.1103/PhysRevLett.38.1440}
  {\path{doi:10.1103/PhysRevLett.38.1440}}.
\newline\urlprefix\url{https://link.aps.org/doi/10.1103/PhysRevLett.38.1440}

\bibitem{kim79}
J.~E. Kim,
  \href{https://link.aps.org/doi/10.1103/PhysRevLett.43.103}{Weak-interaction
  singlet and strong $\mathrm{CP}$ invariance}, Phys. Rev. Lett. 43 (1979)
  103--107.
\newblock \href {http://dx.doi.org/10.1103/PhysRevLett.43.103}
  {\path{doi:10.1103/PhysRevLett.43.103}}.
\newline\urlprefix\url{https://link.aps.org/doi/10.1103/PhysRevLett.43.103}

\bibitem{shifman80}
M.~Shifman, A.~Vainshtein, V.~Zakharov, Can confinement ensure natural cp
  invariance of strong interactions?, Nucl. Phys. B 166 (1980) 493 -- 506.
\newblock \href
  {http://dx.doi.org/https://doi.org/10.1016/0550-3213(80)90209-6}
  {\path{doi:https://doi.org/10.1016/0550-3213(80)90209-6}}.

\bibitem{zhitnitsky80}
A.~R. Zhitnitsky, Sov. J. Nucl. Phys. 31 (1980) 260.

\bibitem{dine81}
M.~Dine, W.~Fischler, M.~Srednicki, A simple solution to the strong {CP}
  problem with a harmless axion, Phys. Lett. B 104 (1981) 199--202.

\bibitem{cast17}
{CAST Collaboration}, \href{http://dx.doi.org/10.1038/nphys4109}{New cast limit
  on the axion–photon interaction}, Nature Phys. 13 (2017) 584.
\newblock \href {http://dx.doi.org/10.1038/nphys4109}
  {\path{doi:10.1038/nphys4109}}.
\newline\urlprefix\url{http://dx.doi.org/10.1038/nphys4109}

\bibitem{alps10}
K.~Ehret, et~al.,
  \href{http://www.sciencedirect.com/science/article/pii/S0370269310005526}{New
  alps results on hidden-sector lightweights}, Phys. Lett. B 689 (2010) 149 --
  155.
\newblock \href
  {http://dx.doi.org/https://doi.org/10.1016/j.physletb.2010.04.066}
  {\path{doi:https://doi.org/10.1016/j.physletb.2010.04.066}}.
\newline\urlprefix\url{http://www.sciencedirect.com/science/article/pii/S0370269310005526}

\bibitem{KEK86}
A.~Konaka, et~al.,
  \href{https://link.aps.org/doi/10.1103/PhysRevLett.57.659}{Search for neutral
  particles in electron-beam-dump experiment}, Phys. Rev. Lett. 57 (1986)
  659--662.
\newblock \href {http://dx.doi.org/10.1103/PhysRevLett.57.659}
  {\path{doi:10.1103/PhysRevLett.57.659}}.
\newline\urlprefix\url{https://link.aps.org/doi/10.1103/PhysRevLett.57.659}

\bibitem{orsay89}
M.~Davier, H.~N. Ngoc,
  \href{http://www.sciencedirect.com/science/article/pii/0370269389901743}{An
  unambiguous search for a light higgs boson}, Phys. Lett. B 229 (1989) 150 --
  155.
\newblock \href
  {http://dx.doi.org/https://doi.org/10.1016/0370-2693(89)90174-3}
  {\path{doi:https://doi.org/10.1016/0370-2693(89)90174-3}}.
\newline\urlprefix\url{http://www.sciencedirect.com/science/article/pii/0370269389901743}

\bibitem{bross91}
A.~Bross, et~al., Phys. Rev. Lett. 67 (1991) 2942.

\bibitem{apex11}
S.~Abrahamyan, et~al.,
  \href{https://link.aps.org/doi/10.1103/PhysRevLett.107.191804}{Search for a
  new gauge boson in electron-nucleus fixed-target scattering by the apex
  experiment}, Phys. Rev. Lett. 107 (2011) 191804.
\newblock \href {http://dx.doi.org/10.1103/PhysRevLett.107.191804}
  {\path{doi:10.1103/PhysRevLett.107.191804}}.
\newline\urlprefix\url{https://link.aps.org/doi/10.1103/PhysRevLett.107.191804}

\bibitem{doebrich16}
B.~D{\"o}brich, J.~Jaeckel, F.~Kahlhoefer, A.~Ringwald, K.~Schmidt-Hoberg,
  \href{https://doi.org/10.1007/JHEP02(2016)018}{Alptraum: Alp production in
  proton beam dump experiments}, JHEP\href
  {http://dx.doi.org/10.1007/JHEP02(2016)018}
  {\path{doi:10.1007/JHEP02(2016)018}}.
\newline\urlprefix\url{https://doi.org/10.1007/JHEP02(2016)018}

\bibitem{kim10b}
J.~E. Kim, G.~Carosi,
  \href{https://link.aps.org/doi/10.1103/RevModPhys.82.557}{Axions and the
  strong $cp$ problem}, Rev. Mod. Phys. 82 (2010) 557--601.
\newblock \href {http://dx.doi.org/10.1103/RevModPhys.82.557}
  {\path{doi:10.1103/RevModPhys.82.557}}.
\newline\urlprefix\url{https://link.aps.org/doi/10.1103/RevModPhys.82.557}

\bibitem{andreas12}
S.~Andreas, C.~Niebuhr, A.~Ringwald,
  \href{https://link.aps.org/doi/10.1103/PhysRevD.86.095019}{New limits on
  hidden photons from past electron beam dumps}, Phys. Rev. D 86 (2012) 095019.
\newblock \href {http://dx.doi.org/10.1103/PhysRevD.86.095019}
  {\path{doi:10.1103/PhysRevD.86.095019}}.
\newline\urlprefix\url{https://link.aps.org/doi/10.1103/PhysRevD.86.095019}

\bibitem{raffelt13}
K.~Baker, et~al., \href{http://dx.doi.org/10.1002/andp.201300727}{The quest for
  axions and other new light particles}, Annalen der Physik 525 (2013)
  A93--A99.
\newblock \href {http://dx.doi.org/10.1002/andp.201300727}
  {\path{doi:10.1002/andp.201300727}}.
\newline\urlprefix\url{http://dx.doi.org/10.1002/andp.201300727}

\bibitem{irastorza18}
I.~G. Irastorza, J.~Redondo, {New experimental approaches in the search for
  axion-like particles}\href {http://arxiv.org/abs/1801.08127}
  {\path{arXiv:1801.08127}}.

\bibitem{mendonca07}
J.~T. Mendonca, Axion excitation by intense laser fields 79 (2007) 21001.

\bibitem{doebrich10}
B.~D{\"o}brich, H.~Gies, Axion-like-particle search using high-intensity
  lasers, JHEP 10 (2010) 1--27.

\bibitem{villalbachavez13}
S.~Villalba-Ch\'avez, {Laser-driven search of axion-like particles including
  vacuum polarization effects}, Nucl. Phys. B881 (2014) 391--413.
\newblock \href {http://arxiv.org/abs/1308.4033} {\path{arXiv:1308.4033}},
  \href {http://dx.doi.org/10.1016/j.nuclphysb.2014.01.021}
  {\path{doi:10.1016/j.nuclphysb.2014.01.021}}.

\bibitem{villalbachavez17}
S.~Villalba-Ch\'avez, T.~Podszus, C.~Müller,
  \href{http://www.sciencedirect.com/science/article/pii/S0370269317302319}{Polarization-operator
  approach to optical signatures of axion-like particles in strong laser
  pulses}, Phys. Lett. B 769 (2017) 233 -- 241.
\newblock \href
  {http://dx.doi.org/https://doi.org/10.1016/j.physletb.2017.03.043}
  {\path{doi:https://doi.org/10.1016/j.physletb.2017.03.043}}.
\newline\urlprefix\url{http://www.sciencedirect.com/science/article/pii/S0370269317302319}

\bibitem{borisov96}
A.~V. Borisov, V.~Y. Grishina, Sov. Phys. JETP 83 (1996) 868--874.

\bibitem{nikishov64}
A.~I. Nikishov, V.~I. Ritus, Quantum processes in the field of a plane
  electromagnetic wave and in a constant field i, Sov. Phys. JETP 19~(2) (1964)
  529--541.

\bibitem{sengupta52}
N.~D. Sengupta, Bull. Math. Soc. 44 (1952) 175.
\newblock \href {http://dx.doi.org/http://dx.doi.org/10.1063/1.1703787}
  {\path{doi:http://dx.doi.org/10.1063/1.1703787}}.

\bibitem{kibble64}
L.~S. Brown, T.~W.~B. Kibble, Interaction of intense laser beams with
  electrons, Phys. Rep. 133~(3A) (1964) A705--A719.

\bibitem{harvey09}
C.~Harvey, T.~Heinzl, A.~Ilderton, Signatures of high-intensity compton
  scattering, Phys. Rev. A 79 (2009) 063407.

\bibitem{mackenroth10}
F.~Mackenroth, A.~Di~Piazza, C.~Keitel, Determining the carrier-envelope phase
  of intense few-cycle laser pulses, Phys. Rev. Lett. 105 (2010) 063903.

\bibitem{khrennikov15}
K.~Khrennikov, J.~Wenz, A.~Buck, J.~Xu, M.~Heigoldt, L.~Veisz, S.~Karsch,
  \href{https://link.aps.org/doi/10.1103/PhysRevLett.114.195003}{Tunable
  all-optical quasimonochromatic thomson x-ray source in the nonlinear regime},
  Phys. Rev. Lett. 114 (2015) 195003.
\newblock \href {http://dx.doi.org/10.1103/PhysRevLett.114.195003}
  {\path{doi:10.1103/PhysRevLett.114.195003}}.
\newline\urlprefix\url{https://link.aps.org/doi/10.1103/PhysRevLett.114.195003}

\bibitem{sakai15}
Y.~Sakai, et~al.,
  \href{https://link.aps.org/doi/10.1103/PhysRevSTAB.18.060702}{Observation of
  redshifting and harmonic radiation in inverse compton scattering}, Phys. Rev.
  ST Accel. Beams 18 (2015) 060702.
\newblock \href {http://dx.doi.org/10.1103/PhysRevSTAB.18.060702}
  {\path{doi:10.1103/PhysRevSTAB.18.060702}}.
\newline\urlprefix\url{https://link.aps.org/doi/10.1103/PhysRevSTAB.18.060702}

\bibitem{yan17}
W.~Yan, et~al.,
  \href{https://www.nature.com/articles/nphoton.2017.100}{High-order
  multiphoton thomson scattering}, Nature Photon. 11 (2017) 514--520.
\newblock \href {http://dx.doi.org/10.1038/nphoton.2017.100}
  {\path{doi:10.1038/nphoton.2017.100}}.
\newline\urlprefix\url{https://www.nature.com/articles/nphoton.2017.100}

\bibitem{reiss62}
H.~R. Reiss, Absorption of light by light, J. Math. Phys. 3~(1) (1962) 59--67.
\newblock \href {http://dx.doi.org/http://dx.doi.org/10.1063/1.1703787}
  {\path{doi:http://dx.doi.org/10.1063/1.1703787}}.

\bibitem{nousch12}
T.~Nousch, D.~Seipt, B.~K\"ampfer, A.~I. Titov, Pair production in short laser
  pulses near threshold, Phys. Lett. B 715 (2012) 246--250.

\bibitem{villalbachavez12}
S.~Villalba-Chavez, C.~Muller, {Photo-production of scalar particles in the
  field of a circularly polarized laser beam}, Phys. Lett. B718 (2013)
  992--997.
\newblock \href {http://arxiv.org/abs/1208.3595} {\path{arXiv:1208.3595}},
  \href {http://dx.doi.org/10.1016/j.physletb.2012.11.035}
  {\path{doi:10.1016/j.physletb.2012.11.035}}.

\bibitem{burke97}
D.~L. Burke, et~al., Positron production in multiphoton light-by-light
  scattering, Phys. Rev. Lett. 79 (1997) 1626.

\bibitem{hu10}
H.~Hu, C.~M{\"u}ller, C.~H. Keitel, Complete qed theory of multiphoton trident
  pair production in strong laser fields, Phys. Rev. Lett. 105 (2010) 080401.

\bibitem{ilderton11}
A.~Ilderton, Trident pair production in strong laser pulses, Phys. Rev. Lett.
  106 (2011) 020404.

\bibitem{king13b}
B.~King, H.~Ruhl, The trident process in a constant crossed field, Phys. Rev. D
  88~(013005) (2013) 013005.

\bibitem{torgrimsson17}
V.~Dinu, G.~Torgrimsson, Trident in plane waves: coherence, exchange and
  space-time inhomogeneity\href {http://arxiv.org/abs/arXiv:1711.04344}
  {\path{arXiv:arXiv:1711.04344}}.

\bibitem{king18a}
B.~King, A.~M. Fedotov, The effect of interference on the trident process in a
  constant crossed field, arXiv preprint arXiv:1801.07300.

\bibitem{ritus85}
V.~I. Ritus, Quantum effects of the interaction of elementary particles with an
  intense electromagnetic field, J. Russ. Laser Res. 6~(5) (1985) 497--617.

\bibitem{dipiazza12}
A.~{Di Piazza}, et~al., Extremely high-intensity laser interactions with
  fundamental quantum systems, Rev. Mod. Phys. 84 (2012) 1177--1228.

\bibitem{narozhny15}
N.~B. Narozhny, A.~M. Fedotov, Extreme light physics, Contemporary Physics 56
  (2015) 249--268.

\bibitem{king15a}
B.~King, T.~Heinzl, {Measuring vacuum polarization with high-power lasers},
  High Power Laser Science and Engineering 4 (2016) e5.
\newblock \href {http://arxiv.org/abs/hep-ph/1510.08456}
  {\path{arXiv:hep-ph/1510.08456}}, \href
  {http://dx.doi.org/doi:10.1017/hpl.2016.1}
  {\path{doi:doi:10.1017/hpl.2016.1}}.

\bibitem{landau4}
V.~B. Berestetskii, E.~M. Lifshitz, L.~P. Pitaevskii, Quantum Electrodynamics
  (second edition), Butterworth-Heinemann, Oxford, 1982.

\bibitem{ilderton09}
T.~Heinzl, A.~Ilderton, A lorentz and gauge invariant measure of laser
  intensity, Opt. Commun. 282 (2009) 1879--1883.

\bibitem{watson22}
G.~N. Watson, Theory of Bessel Functions, Cambridge University Press, Fetter
  Lane, EC4, 1922.

\bibitem{lavelle15}
M.~Lavelle, D.~McMullan,
  \href{https://link.aps.org/doi/10.1103/PhysRevD.91.105022}{Fermionic
  propagator in an intense background}, Phys. Rev. D 91 (2015) 105022.
\newblock \href {http://dx.doi.org/10.1103/PhysRevD.91.105022}
  {\path{doi:10.1103/PhysRevD.91.105022}}.
\newline\urlprefix\url{https://link.aps.org/doi/10.1103/PhysRevD.91.105022}

\bibitem{lavelle17}
M.~Lavelle, D.~McMullan,
  \href{https://link.aps.org/doi/10.1103/PhysRevD.97.036013}{Electrons in an
  eccentric background field}, Phys. Rev. D 97 (2018) 036013.
\newblock \href {http://dx.doi.org/10.1103/PhysRevD.97.036013}
  {\path{doi:10.1103/PhysRevD.97.036013}}.
\newline\urlprefix\url{https://link.aps.org/doi/10.1103/PhysRevD.97.036013}

\bibitem{bamber99}
C.~Bamber, et~al., Phys. Rev. D 60 (1999) 092004.

\bibitem{king16c}
B.~King, H.~Hu,
  \href{http://link.aps.org/doi/10.1103/PhysRevD.94.125010}{Classical and
  quantum dynamics of a charged scalar particle in a background of two
  counterpropagating plane waves}, Phys. Rev. D 94 (2016) 125010.
\newblock \href {http://dx.doi.org/10.1103/PhysRevD.94.125010}
  {\path{doi:10.1103/PhysRevD.94.125010}}.
\newline\urlprefix\url{http://link.aps.org/doi/10.1103/PhysRevD.94.125010}

\bibitem{raicher16}
E.~Raicher, S.~Eliezer, A.~Zigler,
  \href{https://link.aps.org/doi/10.1103/PhysRevA.94.062105}{Nonlinear compton
  scattering in a strong rotating electric field}, Phys. Rev. A 94 (2016)
  062105.
\newblock \href {http://dx.doi.org/10.1103/PhysRevA.94.062105}
  {\path{doi:10.1103/PhysRevA.94.062105}}.
\newline\urlprefix\url{https://link.aps.org/doi/10.1103/PhysRevA.94.062105}

\bibitem{king18b}
B.~M. Dillon, B.~King, (to be submitted)\href {http://arxiv.org/abs/(to be
  submitted)} {\path{arXiv:(to be submitted)}}.

\bibitem{redondo13}
J.~Redondo, \href{http://stacks.iop.org/1475-7516/2013/i=12/a=008}{Solar axion
  flux from the axion-electron coupling}, J. Cosmol. Astropart. P. 12 (2013) 8.
\newline\urlprefix\url{http://stacks.iop.org/1475-7516/2013/i=12/a=008}

\bibitem{mcguffey10}
C.~McGuffey, et~al.,
  \href{https://link.aps.org/doi/10.1103/PhysRevLett.104.025004}{Ionization
  induced trapping in a laser wakefield accelerator}, Phys. Rev. Lett. 104
  (2010) 025004.
\newblock \href {http://dx.doi.org/10.1103/PhysRevLett.104.025004}
  {\path{doi:10.1103/PhysRevLett.104.025004}}.
\newline\urlprefix\url{https://link.aps.org/doi/10.1103/PhysRevLett.104.025004}

\bibitem{esarey09}
E.~Esarey, C.~B. Schroeder, W.~P. Leemans,
  \href{https://link.aps.org/doi/10.1103/RevModPhys.81.1229}{Physics of
  laser-driven plasma-based electron accelerators}, Rev. Mod. Phys. 81 (2009)
  1229--1285.
\newblock \href {http://dx.doi.org/10.1103/RevModPhys.81.1229}
  {\path{doi:10.1103/RevModPhys.81.1229}}.
\newline\urlprefix\url{https://link.aps.org/doi/10.1103/RevModPhys.81.1229}

\bibitem{nakamura17}
K.~Nakamura, et~al., Diagnostics, control and performance parameters for the
  bella high repetition rate petawatt class laser, IEEE Journal of Quantum
  Electronics 53 (2017) 1200121.

\end{thebibliography}

\end{document}